\documentclass{article}
\pdfoutput=1
\usepackage{amssymb}
\usepackage{amsmath}
\usepackage{bm}
\usepackage{commath} % Provides \dif, \norm, \abs, \del, \cbr, \sbr etc.
\usepackage[square,numbers,sort&compress]{natbib} % references

\usepackage{siunitx}
\usepackage{tikz}
\usetikzlibrary{decorations.pathreplacing}
\usepackage{xspace}
\usepackage{authblk} % author affiliations package

\newcommand\matlab{\textsc{Matlab}\textsuperscript{\textregistered}\xspace}
\DeclareMathOperator{\tr}{tr} % trace
\newcommand{\cross}{\pmb{\bm{\times}}} % cross product
\DeclareSIUnit\fps{fps}

\title{Using geometric algebra to represent curvature in shell theory with
  applications to Starling resistors}
\date{31 October 2017}
\author[1,*]{Alastair L Gregory}
\author[1]{Anurag Agarwal}
\author[1]{Joan Lasenby}
\affil[1]{Department of Engineering,
University of Cambridge,
Trumpington Street,
Cambridge, CB2 1PZ}
\affil[*]{\texttt{alg57@cam.ac.uk}}

\begin{document}

\maketitle

\begin{abstract}
  We present a novel application of rotors in geometric algebra to represent the
  change of curvature tensor, that is used in shell theory as part of the
  constitutive law. We introduce a new decomposition of the change of curvature
  tensor, which has explicit terms for changes of curvature due to initial
  curvature combined with strain, and changes in rotation over the surface. We
  use this decomposition to perform a scaling analysis of the relative
  importance of bending and stretching in flexible tubes undergoing self excited
  oscillations. These oscillations have relevance to the lung, in which it is
  believed that they are responsible for wheezing. The new analysis is
  necessitated by the fact that the working fluid is air, compared to water in
  most previous work. We use stereographic imaging to empirically measure the
  relative importance of bending and stretching energy in observed self excited
  oscillations. This enables us to validate our scaling analysis. We show that
  bending energy is dominated by stretching energy, and the scaling analysis
  makes clear that this will remain true for tubes in the airways of the lung.
\end{abstract}

%-------------------------------------------------------------------------------
\section{Introduction}
Self excited oscillations of flexible tubes driven by fluid flow have been a
subject of interest for some time, and there is a considerable literature on the
subject, which is reviewed by
\citep{Heil:2003th,Bertram:2003vq,Grotberg:2004ka,Heil:2011ja,Kisilova:2012us,Jensen:2013td}.
Experimental rigs designed to study this phenomenon are often called Starling
resistors. We are interested in this phenomenon because of its possible
relevance to wheezing in the lung \citep{Grotberg:1980vd}, which is one of the
most commonly heard lung sounds used for diagnosis
\citep{Forgacs:1978ud,Bohadana:2014ei}. Previous work on Starling resistors has
largely used water as the working fluid. In the lung the working fluid is air.
This means that the density ratio between the working fluid and the tube
material in the lung is significantly different from almost all of previously
completed work on Starling resistors. Previous modelling work usually neglects
wall inertia, using instead a ``tube law'' \citep{Whittaker:2010hr}, but there
is strong evidence that wall inertia is significant in the lung from
\citep{Forgacs:1978ud,ShabtaiMusih:1992ty}, who showed that when the density of
the fluid breathed in is changed, the frequency of the wheezes is not affected
significantly. It is clear therefore that the change in density ratio results in
a qualitatively different mechanism. For this reason we have been conducting our
own experiments, and creating models to understand the onset of oscillations.

The flexible tube itself is generally modelled as an elastic shell. Traditional
shell theories
\citep{Leissa:1973tw,Naghdi:1972ul,CiarletJr:2005vn,Antman:1995wm,Lacarbonara:2012bv,Koiter:1966uk}
are well developed but difficult to implement. We have found that linearised
shell theories do not provide good predictions of the frequencies of
oscillation, and we believe that this is due in part to the fact that we have
observed that oscillations start from a collapsed or partially collapsed state.
To use current geometrically non-linear shell theories would require a numerical
simulation of a very complex fluid structure interaction problem, which would be
of similar value to experimental results (though arguably harder to implement),
and would provide the same problem of being difficult to physically interpret
due to the complex nature of the oscillations observed. Instead we would like to
gain a physical understanding of the important mechanisms behind the
oscillations, and this is difficult with shell theories based in differential
geometry \citep{CiarletJr:2005vn,Antman:1995wm}, in particular due to the lack
of physical interpretations of the change of curvature tensor in general
situations. We recently introduced geometric algebra to shell theory
\citep{Gregory:2017dg}, which allowed us to express the fundamental laws in a
component-free form and clarify the role of angular velocity and moments through
the use of bivector representation. For an introduction to the basics of
geometric algebra see \cite{Doran:2003jd}. One of the most powerful aspects of
geometric algebra lies in the use of rotors to represent rotations.  In
\citep{McRobie:1999wv} these have been used to simplify Simo and Vu Quoc's
numerical algorithm \citep{Simo:1988ut} for modelling the nonlinear behaviour of
rods. In projective and conformal geometry \citep[\S10]{Doran:2003jd} rotors
have allowed geometric primitives to be represented in a more simple and lucid
manner, and in relativity \citep{Lasenby:1998gw} and relativistic analogies
\citep{Gregory:2015hc} rotors can simplify transformation between frames of
reference. In this paper we make use of rotors to better understand the change
of curvature of a shell, which is of prime importance to the constitutive law of
the shell, but whose representation has long caused controversy. In
\citep{Leissa:1973tw} at least 10 different linearised shell theories are
presented, and the differences are primarily caused by disagreements over how to
represent changes of curvature.  \citep{CiarletJr:2005vn} has provided a tensor
definition of the change of curvature that has become accepted, however, the
utility of this expression is limited by its complexity. We have been able to
simplify the representation of this tensor using rotors, allowing a more lucid
and physical interpretation of changes of curvature. We take advantage of this
to allow us to understand the importance of the change of curvature in the
context of our Starling resistor experiments.

In order to compare results from the shell theory to our experimental results we
need to be able to calculate the kinematic parameters associated with the
deformation of the flexible tube. To enable this we use stereoscopic imaging,
which to our knowledge is the first time it has been used in the study of
Starling resistors. We take high speed video of the tube at the onset of
oscillation, and are able to track the motion of the surface, and consequently
compare the predictions of shell theory with empirical calculation.

%-------------------------------------------------------------------------------
\section{Understanding Changes in Curvature}
\label{sec:ChangeOfCurvature}
There is an energy associated with any deformation of a shell, and Koiter
\citep{Koiter:1966uk} proposes the following form for this energy,
\begin{equation}
  \label{DeformationEnergy}
  \rho_0U = \frac{Eh}{2(1-\nu^2)}\del{(1-\nu)\tr(\mathsf E^2) +
    \nu\tr(\mathsf E)^2} + \frac{Eh^3}{24(1-\nu^2)}\del{(1-\nu)\tr(\mathsf H^2)
    + \nu\tr(\mathsf H)^2}.
\end{equation}
$U$ is the internal energy per unit mass of the shell, defined on the reference
configuration, $\rho_0$ is the time independent area density of the shell on the
reference configuration, $E$ is Young's modulus, $\nu$ is Poisson's ratio, $h$
is the shell thickness (which in Koiter's theory is assumed constant), $\mathsf
E$ is the two-dimensional Green-Lagrange strain tensor defined on the reference
configuration, $\mathsf H$ is the change of curvature tensor, and $\tr$ is the
trace operator. From \eqref{DeformationEnergy} we can derive the governing
equations of the shell (for more details see \citep{Gregory:2017dg}). The first
term on the right hand side of \eqref{DeformationEnergy} represents the
stretching energy, and the second term represents the bending energy.

In general a shell is a body in which the thickness is smaller than the other
relevant defining length scales. The use of \eqref{DeformationEnergy} for the
energy of deformation implies that we additionally assume that the midsurface of
the body remains the midsurface under deformation, a material line that is
normal to the midsurface remains normal to it under deformation, the shell
thickness remains constant with time, the first and second moments of density
relative to the midsurface are zero, strains within the shell are small, and so
is the normal stress (see \citep{Gregory:2017dg} for further discussion).

Following the notation of \citep{Gregory:2017dg} we take $B$ and $S$ to be the
reference and spatial configurations of the shell, and $X\in B$ and $x\in S$ to
be locations on these configurations. $\phi_t$ is the motion of the shell,
meaning that at time $t$, the point $X\in B$ is at the position $\phi_t(X)\in
S$. $\mathsf G$ and $\mathsf g$ are the identity functions on the reference and
spatial configurations. $Y$ and $y$ are vectors within the tangent spaces of $B$
and $S$ respectively. $\{X^i\},i=1,2$ is a coordinate system on the reference
configuration $B$, which we can then use to define the frame on the reference
configuration $\{E_i=\partial X/\partial X^i\}$. $\{E^i\}$ is the reciprocal
frame that satisfies $E^i\cdot E_j=\delta^i_j$. $\{x^i\}$, $\{e_i\}$ and
$\{e^i\}$ are the similarly defined coordinate system, frame, and reciprocal
frame on the spatial configuration. The shell undergoing deformation is embedded
within a flat three-dimensional Euclidean space $\mathbb E^3$.

If $A$ and $B$ are general multivectors, then $AB$ is the geometric product
between them, $A\cdot B$ is the inner product, $A\wedge B$ is the outer product,
and $A\times B$ is the commutator product, defined by $A\times
B=\frac{1}{2}(AB-BA)$ (see \citep[\S4.1.3]{Doran:2003jd}). We also take $\cross$
(compared to $\times$) to be the cross product between two vectors. If $I$ is
the pseudoscalar of a three-dimensional space, and $a$ and $b$ are vectors, then
$a\cross b=-Ia\wedge b$.

We are particularly interested in the change of curvature that is encoded in
$\mathsf H$. To understand this we must understand the curvature tensors on the
reference and spatial configurations $\mathsf B$ and $\mathsf b$. If $E_3$ and
$e_3$ are the normal vectors to the reference and spatial configurations
respectively, then $\mathsf B(Y)$ and $\mathsf b(y)$ are given by,
\begin{equation}
  \mathsf B(Y) = -Y\cdot\partial E_3,\quad
  \mathsf b(y) = -y\cdot\partial e_3,
\end{equation}
where $\partial$ is the intrinsic vector derivative to any surface. The
relationship between $\partial$ and the vector derivative of $\mathbb E^3$ is
explained in \citep{Gregory:2017dg}.  On $B$ we can expand $\partial$ as
$\partial=E^i\partial/\partial X^i$ and on $S$ we can expand it as
$\partial=e^i\partial/\partial x^i$ \citep[\S6.5.1]{Doran:2003jd}.  $\mathsf B$
and $\mathsf b$ give non-zero results if the surface is not flat.  The change of
curvature tensor $\mathsf H$ is given by,
\begin{equation}
  \mathsf H(Y) = \bar{\mathsf F}\mathsf b\mathsf F(Y) - \mathsf B(Y),
\end{equation}
where $\mathsf F$ is the deformation gradient, defined by $\mathsf
F(Y)=Y\cdot\partial\phi_t(X)$. $\mathsf F$ maps from the tangent space of $B$ to
the tangent space of $S$, providing information about the local deformation of
the surface. $\bar{\mathsf F}$ is the adjoint of $\mathsf F$, i.e. $\mathsf
F(Y)\cdot y=Y\cdot\bar{\mathsf F}(y)$. The strain tensor $\mathsf E$, used in
the constitutive law \eqref{DeformationEnergy}, is given by,
\begin{equation}
  \mathsf E(Y) = \frac{1}{2}\del{\bar{\mathsf F}\mathsf F(Y) - Y}.
\end{equation}
This much is well known, though in other treatments coordinate dependent
definitions of $\mathsf H$ are used (e.g. in \citep{CiarletJr:2005vn}). To make
further progress we will now use rotors to better understand what will produce
changes in $\mathsf H$.

To begin, we note that we can perform a polar decomposition on $\mathsf
F$ such that $\mathsf F(Y)=\mathsf R\mathsf U(Y)$ where $\bar{\mathsf R}=\mathsf
R^{-1}$, $\det\mathsf R=1$, and $\bar{\mathsf U}=\mathsf U$. $\mathsf R$ encodes
rotation, and $\mathsf U$ encodes stretching. We can choose to consider a frame
$\{E_i\}$ on the reference configuration that is locally
orthonormal\footnote{The coordinate system $\{X^i\}$ must be chosen such that
  this is the case, and it may be necessary to use several overlapping
  coordinate systems to achieve this.}. In this case $E_3=E_1\cross
E_2=-I_3(E_1\wedge E_2)=-I_3(E_1E_2)$, where $I_3$ is the pseudoscalar of
three-dimensional Euclidean space $\mathbb E^3$.

To find $e_3$ we need two unit vectors in the tangent space of the spatial
configuration that are oriented in the same way as the pair $\mathsf
F(E_1),\mathsf F(E_2)$, and are orthonormal. This pair of unit vectors is given
by $\mathsf R(E_1),\mathsf R(E_2)$, and so $e_3$ is given by $e_3=\mathsf
R(E_1)\cross\mathsf R(E_2)=-I_3(\mathsf R(E_1)\wedge\mathsf R(E_2))=-I_3(\mathsf
R(E_1)\mathsf R(E_2))$. The fact that the function $\mathsf R$ is a rotation
means that it has an associated rotor $R$ such that $\mathsf R(Y)=RY\!\tilde R$,
where $R$ is an even multivector that satisfies $\tilde RR=R\tilde R=1$, where
$\tilde R$ is the reverse of $R$. This allows us to write $e_3$ as,
\begin{equation}
  e_3 = -I_3\del{(RE_1\tilde R)(RE_2\tilde R)} =
  -I_3\del{R(E_1E_2)\tilde R} = RE_3\tilde R,
\end{equation}
where we have used the fact that any rotor will commute with $I_3$. Thus we have
shown that the rotation associated with the deformation is also the rotation
between the normal vectors $E_3$ and $e_3$, which makes intuitive sense. We have
also extended the range and domain of $\mathsf R$ to $\mathbb E^3$, while the
range and domain of $\mathsf U$ is still constrained to the tangent space of the
reference configuration, and the range and domain of $\mathsf F$ are constrained
to the tangent spaces of the spatial and reference configuration respectively.

Two results that we will find useful are,
\begin{subequations}
\begin{gather}
  \label{RotorReverseDerivative}
  Y\cdot\partial\tilde R = -\tilde R(Y\cdot\partial R)\tilde R, \\
  \label{TransportVectorDerivative}
  \mathsf F(Y)\cdot\partial e_3 = Y\cdot\partial e_3.
\end{gather}
\end{subequations}
\eqref{RotorReverseDerivative} follows from $\tilde RR=1$.
\eqref{TransportVectorDerivative} has implicit assumptions that require
explanation. The expression on the left of \eqref{TransportVectorDerivative}
tells us how $e_3$ varies over the spatial configuration in the direction
defined by $\mathsf F(Y)$, which lies in the tangent space of the spatial
configuration. On the right of \eqref{TransportVectorDerivative}, $e_3=e_3(x)$
has been mapped to a vector field on the reference configuration such that
$e_3(X)=e_3(\phi_t(X))\;\forall X\in B$. This allows the expression on the right
to tell us how $e_3$ varies over the reference configuration in the direction
defined by $Y$, which is tangent to the reference configuration. The equality of
these expressions is a standard result when mapping derivatives between
manifolds, which can be proven by considering derivatives with respect to
convected coordinates that satisfy $x^i(x) = X^i(\phi_t^{-1}(x))$. From this
point we will assume that $\{x^i\}$ are convected coordinates.

Using these results we can write $\bar{\mathsf F}\mathsf b\mathsf F(Y)$ as
$\bar{\mathsf F}\mathsf b\mathsf F(Y)=-\bar{\mathsf F}(Y\cdot\partial e_3)$, and
the argument of $\bar{\mathsf F}$ can be expressed as,
\begin{equation}
  \begin{aligned}
    Y\cdot\partial e_3 = Y\cdot\partial(RE_3\tilde R) &=
    (Y\cdot\partial R)E_3\tilde R + R(Y\cdot\partial E_3)\tilde R +
    RE_3(-\tilde R(Y\cdot\partial R)\tilde R) \\
    &= R(Y\cdot\partial E_3)\tilde R +
    [(Y\cdot\partial R)\tilde R]RE_3\tilde R -
    RE_3\tilde R[(Y\cdot\partial R)\tilde R] \\
    &= R(Y\cdot\partial E_3)\tilde R +
    [(Y\cdot\partial R)\tilde R]e_3 -
    e_3[(Y\cdot\partial R)\tilde R] \\
    &= R(Y\cdot\partial E_3)\tilde R +
    [2(Y\cdot\partial R)\tilde R]\times e_3 \\
    &= \mathsf R(Y\cdot\partial E_3) +
    [2(Y\cdot\partial R)\tilde R]\times e_3,
  \end{aligned}
\end{equation}
where $\times$ is the commutator product. Hence we can express $\bar{\mathsf
  F}\mathsf b\mathsf F(Y)$ as,
\begin{equation}
  \begin{aligned}
    \bar{\mathsf F}\mathsf b\mathsf F(Y) &=
    -\bar{\mathsf F}\mathsf R(Y\cdot\partial E_3) -
    \bar{\mathsf F}\del{[2(Y\cdot\partial R)\tilde R]\times e_3} \\
    &= - \mathsf U(Y\cdot\partial E_3) -
    \bar{\mathsf F}\del{[2(Y\cdot\partial R)\tilde R]\times e_3} \\
    &= \mathsf U\mathsf B(Y) +
    \bar{\mathsf F}\del{e_3\times[2(Y\cdot\partial R)\tilde R]},
  \end{aligned}
\end{equation}
and finally we obtain an expression for $\mathsf H$,
\begin{equation}
  \begin{aligned}
    \mathsf H(Y) &=
    (\mathsf U-\mathsf G)\mathsf B(Y) +
    \bar{\mathsf F}\del{e_3\times[2(Y\cdot\partial R)\tilde R]} \\
    &= (\mathsf U-\mathsf G)\mathsf B(Y) +
    \bar{\mathsf F}\del{[RE_3\tilde R]\times[2(Y\cdot\partial R)\tilde R]}.
  \end{aligned}
\end{equation}
This shows that there are two contributions to $\mathsf H$. Firstly, if the
reference configuration is at all curved (i.e. $\mathsf B(Y)$ is non-zero), then
the strain of the shell, encoded in $\mathsf U-\mathsf G$, will result in a
change of curvature. The second contribution is due to variation of the rotor
$R$ over the shell. These two kinds of change of curvature are illustrated well
by an inflating sphere and deformation of a flat plate. As a sphere is inflated
to become a larger sphere, the normal vector is unchanged, i.e. $e_3=E_3$, hence
$R=1$ everywhere. This means that the second term in our expression for $\mathsf
H$ will be zero. However, the surface of the sphere will stretch, meaning that
$\mathsf U-\mathsf G$ will be non-zero. In addition, $\mathsf B$ will be
non-zero for a sphere, which tells us that the first term in our expression for
$\mathsf H$ will be non-zero. By contrast, for a flat plate $\mathsf B(Y)$ will
be zero, meaning that $R$ must vary over the plate in order for there to be any
change of curvature.

A variant on this expression for $\mathsf H$ can be obtained if we express the
rotor $R$ as $R=\exp(-A/2)=\exp(-\hat A\theta/2)$, where $A$ is a bivector
aligned with the plane of rotation whose magnitude is equal to the angle of
rotation $\theta$. $\hat A$ is a unit bivector ($\hat A^2=-1$). Note that the
direction of rotation is defined by the sign of $\theta$ and the orientation of
$\hat A$ together. We can set the convention that $\theta\geqslant0$, in which
case $\theta$ and $\hat A$ are uniquely defined if $A$ is known. Given this
definition we can express $Y\cdot\partial R$ as,
\begin{equation}
  Y\cdot\partial R = -\frac{Y\cdot\partial A}{2}\exp(-A/2) =
  -\frac{Y\cdot\partial A}{2}R.
\end{equation}
Using this we can express $\mathsf H$ as,
\begin{equation}
  \mathsf H(Y) = (\mathsf U-\mathsf G)\mathsf B(Y) -
  \bar{\mathsf F}\del{(RE_3\tilde R)\times(Y\cdot\partial A)}.
\end{equation}
The term that $\bar{\mathsf F}$ operates on is the commutator product of a
vector and bivector, so we can replace the commutator product with a dot
product,
\begin{equation}
  \label{HRepresentation}
  \mathsf H(Y) = (\mathsf U-\mathsf G)\mathsf B(Y) -
  \bar{\mathsf F}\del{(RE_3\tilde R)\cdot(Y\cdot\partial A)}.
\end{equation}
Taking the inner product of a bivector with $RE_3\tilde R=e_3$ means that only
vectors tangential to the spatial configuration are retained, which then means
that $\bar{\mathsf F}$ can operate and return vectors tangential to the
reference configuration. Hence our description confirms that the range and
domain of $\mathsf H$ are both the tangent space of the reference configuration.

The explicit expression for the two possible contributions to change of
curvature, shown in \eqref{HRepresentation}, gives a new decomposition of the
change of curvature tensor which will be of use when we try to understand the
importance of bending in the Starling resistor.

%-------------------------------------------------------------------------------
\section{Experiment Description}
%-------------------------------------------------------------------------------
\subsection{Experimental Setup}
\begin{figure}
  \centering
  \includegraphics{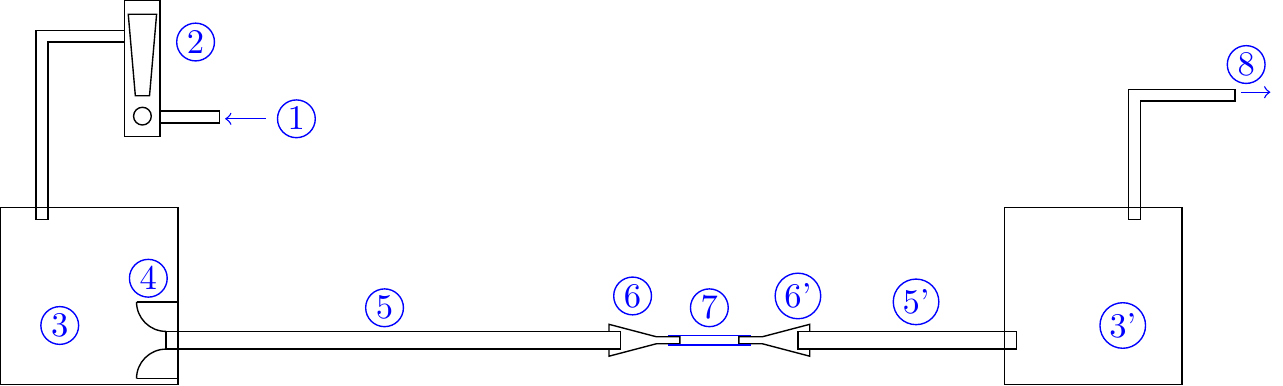}
  \caption{Schematic of Starling resistor experiment. 1: Flow inlet, 2:
    Rotameter, 3/3': Settling chambers, 4: Clean flow inlet, 5/5': Clean flow
    tubes, 6/6': Contraction and expansion, 7: flexible tube, 8: tube to suction
    fan. The downstream settling chamber is approximately \SI{4}{\cubic\m} while
    the upstream settling chamber is \SI{0.03}{\cubic\m}.}
  \label{ExperimentSchematic}
\end{figure}

\figref{ExperimentSchematic} shows a schematic of the experimental setup used to
investigate the oscillations of flexible tubes. Air flows into the system
through (1), then through a rotameter (2) used to monitor flowrate. The noise
that the rotameter introduces into the flow, and any other noise, is isolated
from the flexible tube by the upstream settling chamber (3). Air flows into the
upstream clean flow tube (5) section via a shaped inlet (4) that reduces
separation. A contraction (6) leads to the flexible tube (7), before an
expansion (6') leads to the downstream clean flow tube (5') that exits into the
downstream settling chamber (3'). Suction is provided by a fan (8). The
downstream settling chamber (3') isolates the flexible tube from the noise from
this fan. Experiments were performed in the Acoustics Laboratory in the
Department of Engineering at the University of Cambridge.

In the experiments relevant to this paper the suction at (8) is gradually
increased until the flexible tube just starts oscillating. With the tube
oscillating in this way high speed video is recorded from 2 FASTCAM-ultima APX
cameras (produced by Photron \texttt{https://photron.com}) with a frame rate of
\SI{12500}{\fps} and a resolution of $512\times256$ pixels (grayscale). Our
experiments require us to focus on a small flexible tube at reasonably close
range. The Photron camera has an adaptor for Nikon lens's. We use a \SI{50}{\mm}
lens combined with a \SI{7}{\mm} extension tube to allow us to focus on the tube
and have it fill most of the frame. An aperture of f/2.8 is used.

The flexible tubes used are made out of rubber latex for which $E=\SI{1}{\MPa}$
and $\nu=\num{0.5}$. The tube diameter is \SI{6}{\mm}, the wall thickness is
\SI{0.3}{\mm} and the unstrained length is \SI{19}{\mm}. The tubes are held in
an axially strained state, so the length of the tubes during the experiment is
\SI{25}{\mm}.

%-------------------------------------------------------------------------------
\subsection{Image Processing}
\label{sec:imageProcessing}
\begin{figure}
  \centering
  \includegraphics{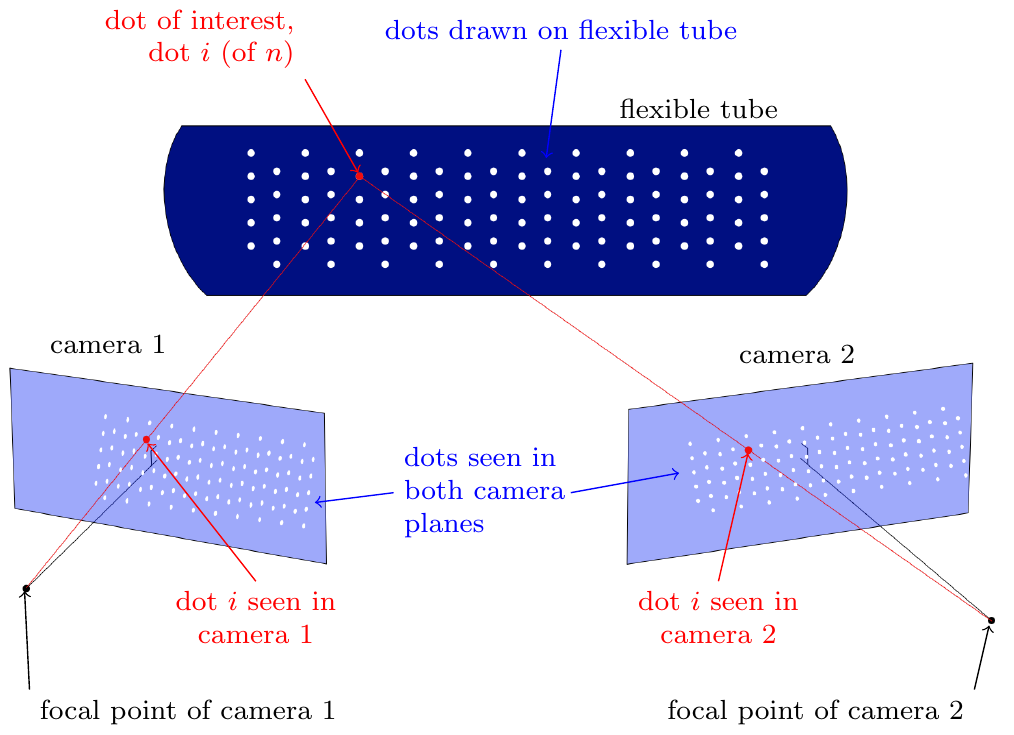}
  \caption{Schematic of the high speed camera setup.}
  \label{HSCschematic}
\end{figure}

The high speed cameras record at \SI{12500}{\fps}, and are triggered together,
so that every $\Delta t=\SI{80}{\micro\s}$ two images of the flexible tube are
taken. A schematic of the two cameras and the flexible tube is shown in
\figref{HSCschematic}. Dots are drawn on the flexible tube (shown in white in
\figref{HSCschematic}), which indicate a set of material points we would like to
track over time in three dimensions.

It is possible to find the characteristics\footnote{The characteristics of an
  imaginary pinhole camera that we replace the real camera with to allow us to
  use methods from projective geometry to perform the triangulation.} of two
cameras such that if a point appears in simultaneous images from both cameras,
the point's position in three-dimensional space can be triangulated
\citep{Dorst:2009tt,Hartley:2004un}. Calibration involves taking at least
\num{3}, and in general between \num{10} and \num{20}, simultaneous images of a
chequerboard pattern in various orientations. From this the position of the two
cameras relative to each other, and their internal parameters, can be
calculated. In this method cameras are modelled as pinhole cameras, meaning that
the focal length, pixel size and skew are the important internal parameters.  In
addition it is possible to account for radial distortion of the image by the
camera lens, and tangential distortion, which occurs when the image sensor is
not perfectly perpendicular to the line of sight of the camera. This calibration
is performed using the computer vision system toolbox of \matlab
\citep{MATLAB:2017uk}. The images produced by these pinhole camera models is
what is illustrated in \figref{HSCschematic}.

To find the three-dimensional tracks of the material points, we must first
find the locations of the dots within each image. We refer to these coordinate
pairs as points. The dots on the tube surface are drawn in white, are spaced by
approximately \SI{1}{\mm}, and have a diameter of approximately \SI{0.7}{\mm}.
We take the material points to be the centres of the dots, and we find them by
first taking a two-dimensional convolution of the image with a ``mexican hat''
function of the form,
\begin{equation}
  \frac{1}{\pi\sigma^4}\del{1-\frac{x^2+y^2}{2\sigma^2}}
  e^{-\frac{x^2+y^2}{2\sigma^2}},
\end{equation}
where $\sigma$ is the expected radius of the dot in the image. The convolution
effectively smooths the image, removing any artefacts from drawing the dots,
leaving peaks in the centroid of each dot. These peaks are then used as the
locations of each point.

Hence at each time instance, we have two collections of points, representing the
material points as seen from each camera. If there are $n$ points on the tube,
and $m$ frames in our video, then in total we will have found the locations of
$2nm$ points. To make use of this data, we need to identify each unique material
point in each camera and over time.

To associate points across time for a single camera's set of images, the points
at $t$ and $t+\Delta t$ are compared, and if two points are within a certain
distance of each other, then it is assumed that these represent the same point.
This works because the frame rate (\SI{12500}{\fps}) is much larger than the
frequency of the observed vibrations ($\sim\SI{500}{\Hz}$), so motions between
frames are small.

A pair of corresponding points in the two camera images are illustrated in
\figref{HSCschematic}, but finding these pairings at each instant in time is
more complex. First we consider the line drawn from the focal point of camera 2
to dot $i$ in the image, which we will call a ray. Anything on this ray in
three-dimensional space will appear at the same highlighted location in camera
2. However, from camera 1, the ray will appear as a line. Therefore, if a
point's location is known in one camera image, then it must lie on a specific
line in the other image. This line is known as an epipolar line
\citep{Hartley:2004un}. Hence, for a pair of points, one in each camera image,
to correspond to the same material point, they must each lie on the epipolar
line of the other. However, because of the specific arrangement of the cameras
and dots, this does not usually provide a unique set of pairs. The relative
positioning of the cameras means that the epipolar lines are all approximately
horizontal, and the dots drawn on the tube are arranged in horizontal rows, so
multiple dots can be very close to a given epipolar line. To overcome this, we
specify \num{10} corresponding point pairs between the two images (\num{20}
points in total), and these \num{10} pairs are then used to find the best
fitting projective transformation from camera 1 to camera 2. Applying this
transformation to the image from camera 1 places each point close to the
corresponding point in the image from camera 2, allowing all the remaining point
pairs to be found. This result is then checked for consistency with the epipolar
line condition.

In addition to tracing material points over time as the self excited
oscillations occur, we must also find the locations of the material points on
the tube when it is unstrained. This is necessary for the calculation of the
kinematic variables used in expression for energy of deformation
\eqref{DeformationEnergy}. To achieve this a single image is taken from each
camera when the tube is held in its unstrained state. Pairing of points must
then also be completed between the two images of the tube in its unstrained
state and images from the high speed video of self excited oscillations. This
pairing is done using the methods described in the previous paragraph.

Once point pairs are known over time, the camera calibration can be used to find
three-dimensional point traces over time. The spatial resolution of this trace
is limited by the size of the pixels in the high speed video. This results in
point traces with distinct jumps in position. These jumps are by no more than
\SI{0.1}{\mm} in three-dimensional space, compared to variations in position
on the order of \SI{2}{\mm} over the course of the self excited oscillations.
For this reason we smooth the three-dimensional point traces by fitting
functions of the form,
\begin{equation}
  \sum_{i=1}^8 A_i\sin(\omega_i t),
\end{equation}
to the three position components, where $A_i$ and $\omega_i$ are chosen to fit
the empirical data. These fits work well because the videos are of quasi-steady
behaviour at the onset of oscillation, and the observed motions are close to
sinusoidal. \num{8} terms have been found to be sufficient to match the
experimental data. This smoothing is necessary in order for derivatives of the
point traces to give meaningful results.

%-------------------------------------------------------------------------------
\subsection{Kinematic Calculations}
\begin{figure}
  \centering
  \includegraphics[width=13cm]{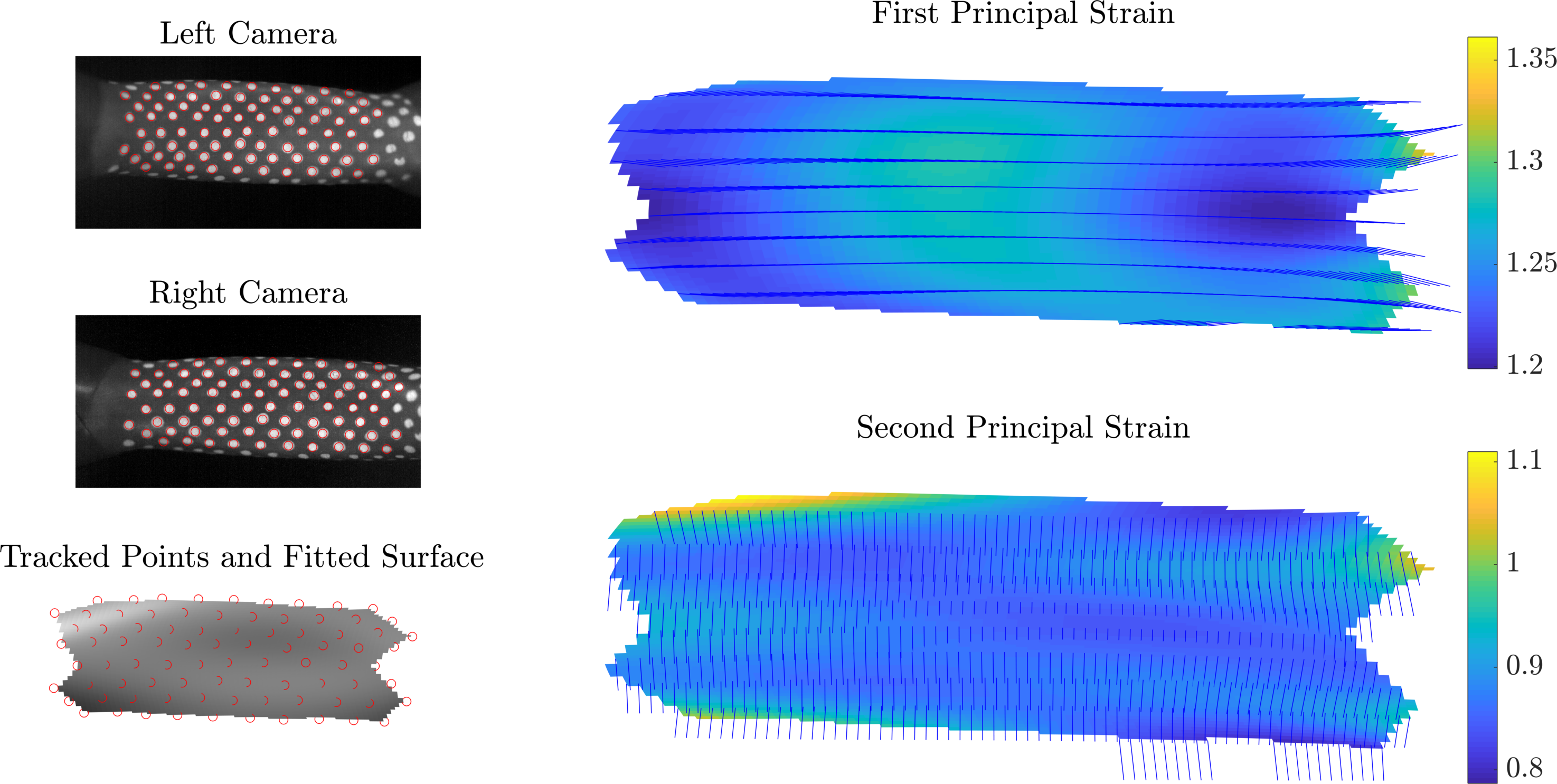}
  \caption{A typical image of the principal strains of the flexible tubes, shown
    along with the principal strain directions (illustrated with unit vectors).
    On the left the original high speed camera images are shown with the tracked
    surface points, and a view of the three-dimensional triangulation with the
    surface fitted to them.}
  \label{HighSpeedVideoStill_Strain}
\end{figure}

In \figref{HighSpeedVideoStill_Strain} we show a single frame from the video in
which the dots on the surface of the tube have been automatically detected,
identified with the corresponding dots in the other video image, and identified
with the corresponding dots in a stereoscopic image of the unstrained tube (not
shown). With this information we can reconstruct the points in three
dimensions and fit a surface to them. The process can be summarised as follows,
\begin{itemize}
  \item Locate points in images of unstrained tube.
  \item Track points in high speed video frames.
  \item Associate points between all images (as described in
    \secref{sec:imageProcessing}) and triangulate to have the position of each
    material point as a function of time, and its position on the unstrained
    tube.
  \item Assign a pair of coordinate values $\{x^i\}$ to each point for use as
    the convected surface coordinates.
  \item Fit a smoothing curve to every three-dimensional point as a function
    of time as described in \secref{sec:imageProcessing}.
  \item At a chosen time take the positions of all of the points and fit a
    polynomial surface such that we have position as a function of the surface
    coordinates $\{x^i\}$. Repeat this for the unstrained surface.
  \item Take all possible first and second derivatives of the surface position
    with respect to $\{x^i\}$. This is done analytically using the polynomial
    surface. Repeat this for the unstrained surface.
\end{itemize}
The surface fit and its derivatives are used to calculate all the kinematic
properties of the undeformed and deformed surfaces.

\begin{figure}
  \centering
  \includegraphics[width=13cm]{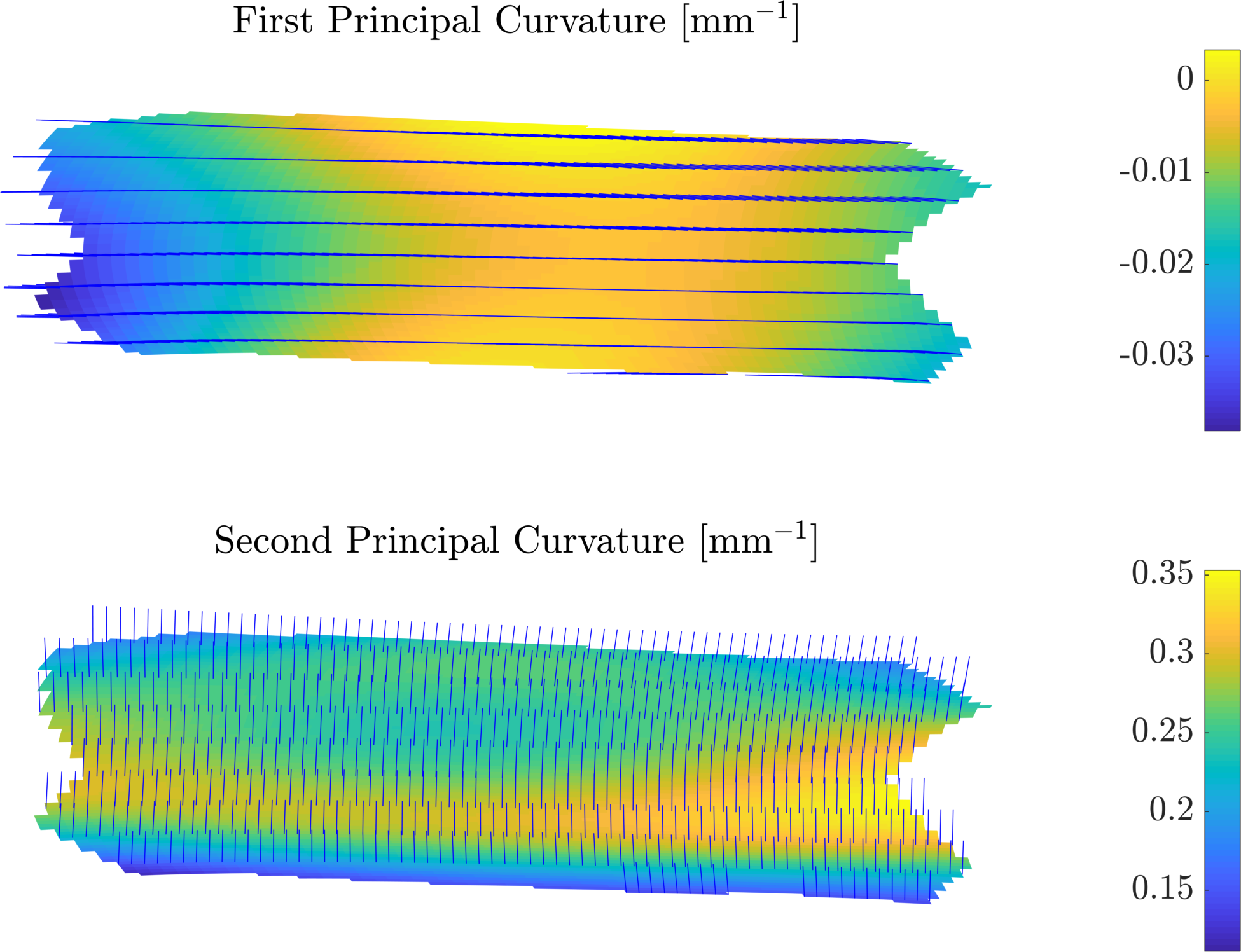}
  \caption{A typical image of the principal curvatures of the flexible tube,
    shown along with the principal curvature directions (illustrated with unit
    vectors).}
  \label{HighSpeedVideoStill_Curvature}
\end{figure}

In \figref{HighSpeedVideoStill_Strain} we show a typical image of the principal
strains of the flexible tube, i.e. the eigenvectors and values of $\mathsf E$.
An eigenvalue of $1$ corresponds to no strain, a value less than \num{1}
corresponds to compression, and a value greater than \num{1} corresponds to
tension. The eigenvectors give the direction in which the strain is occurring.
We can see that the first principal strain is approximately aligned with the
longitudinal direction, and is tensile. This is due to the dominant pre-strain
of the elastic tube. By contrast, the second principal strain, which is
primarily in the azimuthal direction, is a mixture of compression and tension,
and is generally closer to \num{1}. In \figref{HighSpeedVideoStill_Curvature} we
show the principal curvatures of the deformed surface, i.e. the eigenvectors and
values of $\mathsf b$. For a cylinder, which the tube is in its undeformed
state, the principal curvatures would be \num{0} in the longitudinal direction
and $1/a=\SI{0.33}{\per\mm}$ in the azimuthal direction. We see that in the
deformed tube the curvatures are still closely aligned to the longitudinal and
azimuthal directions, and can see that the squashing of the tube results in a
slightly negative longitudinal curvature and a slight reduction in the azimuthal
curvature towards the centre of the tube. But these effects are fairly small.

%-------------------------------------------------------------------------------
\section{Calculation of Bending and Stretching Energies}
We now aim to gain more understanding of how the tube deforms. To do this we
will use a mixture of empirical and analytical techniques. More specifically we
can use the high speed video reconstructions combined with the mathematical
framework for shells already developed in \secref{sec:ChangeOfCurvature}.

%-------------------------------------------------------------------------------
\subsection{Scaling Analysis}
We start by estimating the bending and stretching energies analytically, which
requires us to estimate the values of $\tr(\mathsf E^2)$, $\tr(\mathsf E)^2$,
$\tr(\mathsf H^2)$, and $\tr(\mathsf H)^2$.

If $\alpha_i$ are the eigenvalues of the linear function $\mathsf A$, then
$\tr(\mathsf A)^2=\alpha_1^2+2\alpha_1\alpha_2+\alpha_2^2$ and $\tr(\mathsf
A^2)=\alpha_1^2+\alpha_2^2$. We know that the eigenvalues of $\mathsf E$ are
$\frac{1}{2}(\lambda_i^2-1)$ ($\lambda_i$ are the principal strains) and that
$\lambda_1\sim\lambda$, $\lambda_2\sim1$, where $\lambda$ is the initial axial
strain of the tube. This allows us to get an order of magnitude estimate for
$\tr(\mathsf E^2)$ and $\tr(\mathsf E)^2$ of $(\lambda^2-1)^2/4$. If we take
$\lambda=1.3$\footnote{The strained length of the tube is $l=\SI{25}{\mm}$ and
  the unstrained length is $l_0=\SI{19}{\mm}$, giving
  $\lambda=l/l_0\approx1.3$.} then we have $\tr(\mathsf E^2)\sim\tr(\mathsf
E)^2\sim\num{0.1}$.

To understand bending in the tube we consider the representation of $\mathsf H$
derived in \secref{sec:ChangeOfCurvature}, and given in \eqref{HRepresentation}.
We can write the action of $\mathsf U(Y)$ as $\mathsf U(Y)=\lambda_1(Y\cdot\hat
W_1)\hat W_1+\lambda_2(Y\cdot\hat W_2)\hat W_2$, where $\hat W_i$ are the unit
eigenvectors of $\mathsf U$. Using the approximation $\lambda_1\approx\lambda$
and $\lambda_2\approx1$ this becomes $\mathsf U(Y)\approx\lambda(Y\cdot\hat
W_1)\hat W_1+(Y\cdot\hat W_2)\hat W_2$. We can write $\mathsf B(Y)$ as $\mathsf
B(Y)=C(Y\cdot\hat E_2)\hat E_2$, where $\hat E_i$ are the unit eigenvectors of
$\mathsf B$, and $C$ is the principal curvature of the undeformed tube in the
azimuthal direction. Combining these we have,
\begin{equation}
  \begin{aligned}
    \mathsf U\mathsf B(Y) - \mathsf B(Y) &= \mathsf U(CY\cdot\hat E_2\hat E_2) -
    CY\cdot\hat E_2\hat E_2 \\
    &\approx CY\cdot\hat E_2(\lambda\hat E_2\cdot\hat W_1\hat W_1 +
    \hat E_2\cdot\hat W_2\hat W_2) - CY\cdot\hat E_2\hat E_2.
  \end{aligned}
\end{equation}
We know from \figref{HighSpeedVideoStill_Strain} that $\hat W_1$ and $\mathsf
W_2$ are approximately aligned with the longitudinal and circumferential
directions, so we can write $\hat E_2\cdot\hat W_1\approx0$ and $\hat
E_2\cdot\hat W_2\approx1$. Using this we obtain,
\begin{equation}
  \mathsf U\mathsf B(Y) - \mathsf B(Y) \approx
  CY\cdot\hat E_2\hat W_2 - CY\cdot\hat E_2\hat E_2 \approx 0.
\end{equation}
This is saying that because the directions of principal strain and principal
curvature are approximately perpendicular, the influence of strain on change of
curvature is removed.

\begin{figure}
  \centering
  \includegraphics[width=12cm]{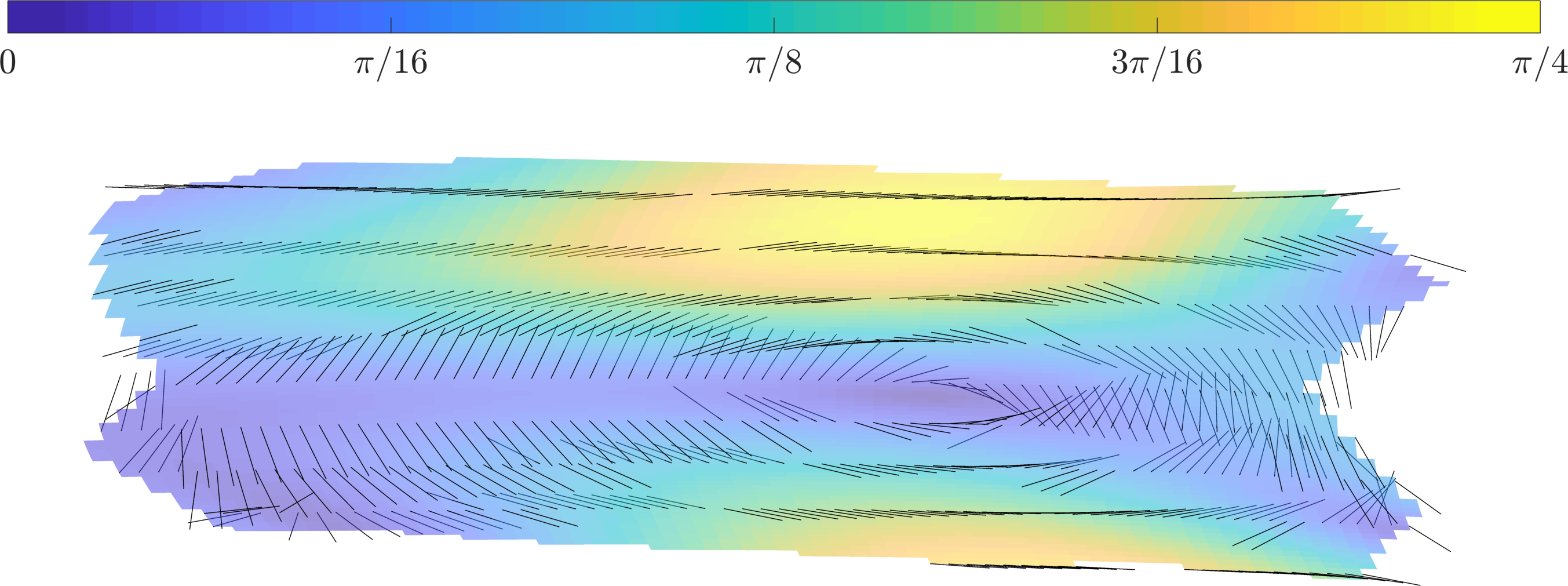}
  \caption{A visualisation of the rotation $\mathsf R$. The axis of rotation is
    shown along with the absolute value of the angle of rotation (in radians).}
  \label{HSV_rotation}
\end{figure}

We are now in a position to consider the rotor $R$, since it is changes in this
multivector over the surface of the shell that are responsible for the change of
curvature. $R$ represents the rotation, and as is shown in
\secref{sec:ChangeOfCurvature} is characterised by the bivector $A=\theta\hat
A$, whose magnitude gives the rotation angle in radians, and whose plane gives
the plane of rotation. In \figref{HSV_rotation} we visualise the angle and axis
of rotation encoded in the rotation tensor $\mathsf R$, corresponding to a
typical deformation. The axes of rotation shown in \figref{HSV_rotation} are
primarily tangential to the surface, so the bivector $A$ will be dominated by
the components $e_1\wedge e_3$ and $e_2\wedge e_3$, with little rotation in the
$e_1\wedge e_2$ plane, i.e. about the normal vector $e_3$. Hence, we can write
$A$ as,
\begin{equation}
  A = \theta_1 e^1\wedge e_3 + \theta_2 e^2\wedge e_3 = \theta_i e^i\wedge e_3.
\end{equation}
We have used the reciprocal frame $\{e^i\}$ instead of of $\{e_i\}$ because it
will allow us to use the property $\bar{\mathsf F}(e^i)=E^i$\footnote{The
  coordinate system is convected, so $e_i=\mathsf F(E_i)$, and as always
  $e^i\cdot e_j=E^i\cdot E_j=\delta^i_j$. Hence $\delta^i_j=e^i\cdot\mathsf
  F(E_j)=\bar{\mathsf F}(e^i)\cdot E_j$, from which it is clear that
  $\bar{\mathsf F}(e^i)=E^i$.}. We can extend the
frames $\{e^i\}$ and $\{e_i\}$ to span $\mathbb E^3$ by using the normal vector
$e_3$. Because $e_3$ is a unit vector and perpendicular to all of $e_i,e^i$, we
can also write $e^3=e_3$, and we have the frame $\{e_a\},a=1,2,3$ and $\{e^a\}$.
Using this we define the Christoffel coefficients
$\gamma^a_{ib}=e^a\cdot\partial e_b/\partial x^i$, $i=1,2;a,b=1,2,3$. These also
satisfy $e_b\cdot\partial e^a/\partial x^i=-\gamma^a_{ib}$.

Substituting $A$ into the second part of the change of curvature tensor given in
\eqref{HRepresentation}, using the fact that $e_3$ is normal to $e^1$ and $e^2$,
and $\gamma^3_{i3}=0$, we obtain,
\begin{equation}
  \begin{aligned}
    \bar{\mathsf F}\del{e_3\cdot\del{Y\cdot\partial A}} &=
    Y^i\bar{\mathsf F}\del{-\partial_i(\theta_j)e^j+\theta_j\gamma^j_{ik}e^k} \\
    &= -Y^i\partial_i(\theta_j)E^j + Y^i\theta_j\gamma^j_{ik}E^k.
  \end{aligned}
\end{equation}
Therefore, given that the first part of $\mathsf H$ in \eqref{HRepresentation}
is zero, $E_i\cdot\mathsf H(E_j)=\mathsf H_{ij}$ is given by,
\begin{equation}
  \mathsf H_{ij} = \partial_j\theta_i - \theta_k\gamma^k_{ji}.
\end{equation}
We know that $\mathsf H$ is symmetric, so from this we see that our earlier
assumption on the form of $A$ must be joined by the condition
$\partial_i\theta_j=\partial_j\theta_i$ to produce consistent results.

\begin{figure}
  \centering
  \includegraphics{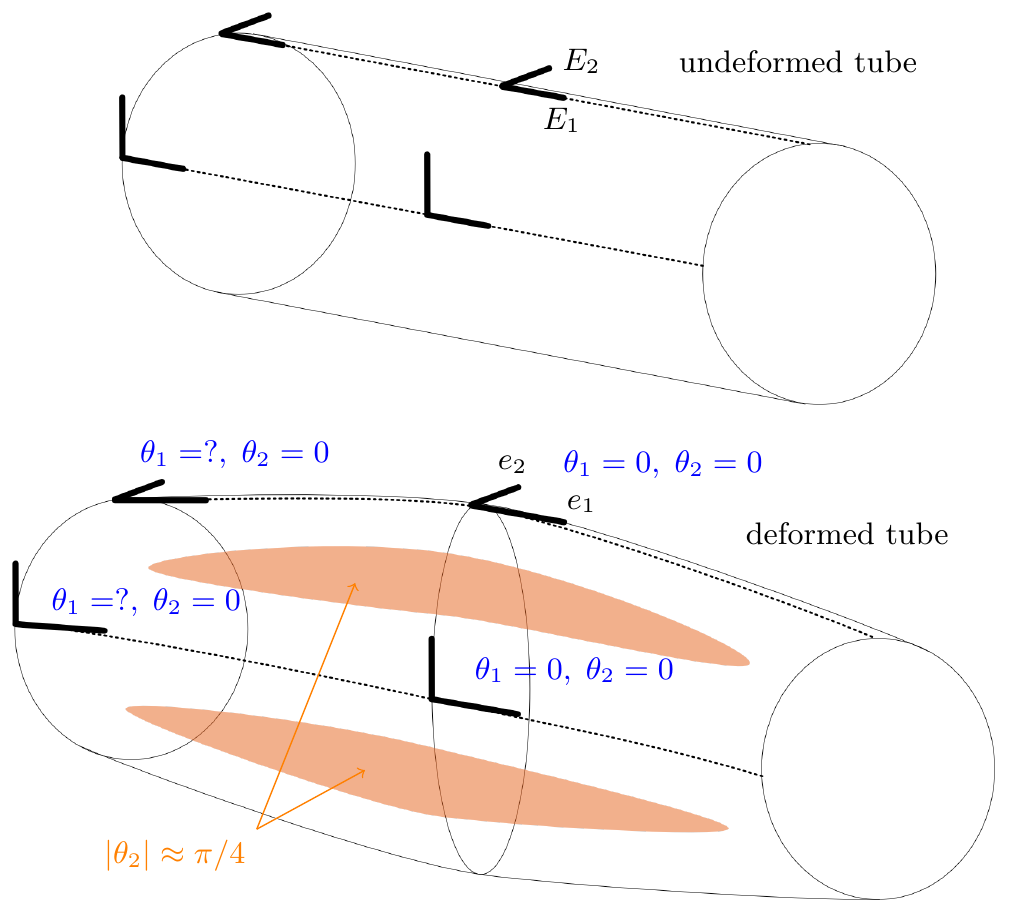}
  \caption{A schematic of the deformation of the flexible tube.}
  \label{BendingSchematic}
\end{figure}

The frame $\{E_i\}$ can be chosen by us to be orthonormal. More specifically, we
can align $E_1$ with the longitudinal direction and $E_2$ with the azimuthal
direction on the unstrained cylindrical tube. From \figref{HighSpeedVideoStill_Strain} we can see that
under a typical deformation these basis vectors remain close to the axial and
azimuthal directions. In \figref{BendingSchematic} we give a schematic
illustration of how $E_i$ maps to $e_i$. In \figref{BendingSchematic}
we have also labelled the values of $\theta_i$ where they are obvious. The
regions where $\abs{\theta_2}\approx\pi/4$ can be seen empirically in
\figref{HSV_rotation}. There are two unknown values of $\theta_1$ shown as
question marks. We can estimate the largest value of these rotations by assuming
a straight line from the clamped tube end and the centre of the tube when the
tube collapses completely at the centre. In this case
$\theta_1\sim\arctan(a/(l/2))\lambda$ where $a$ is the tube radius and $l$ is
the tube length in its deformed state. The multiplication by $\lambda$ is
necessary because $\theta_1$ is the $e^1\wedge e_3$ component, and $e^1$ is
shortened by a factor of $\lambda$ compared to the unit vector $E^1$. Up to
angles of \SI{30}{\degree}, $\tan\theta$ is within \SI{10}{\percent} of $\theta$,
so we will take $\theta_1\sim \lambda a/(l/2)=a/(l_0/2)$ at the point in
question, where $l_0$ is the unstrained length of the tube. This, and the values
of $\theta_i$ shown in \figref{BendingSchematic}, allows us to make the
following estimates,
\begin{equation}
  \begin{aligned}
    \partial_1\theta_1 &\sim \frac{a}{l_0/2}\frac{1}{l_0/2} = \frac{4a}{l_0^2}, \\
    \partial_1\theta_2 = \partial_2\theta_1 &\sim \frac{\pi/4}{l_0/2} =
    \frac{\pi}{2l_0}, \\
    \partial_2\theta_2 &\sim \frac{\pi/4}{2\pi a/8} = \frac{1}{a}.
  \end{aligned}
\end{equation}
Because of replacement of $\arctan$ with the identity function, our estimate for
$\partial_1\theta_1$ will be an overestimate when the tube is very short.

We can also estimate the values of the coefficients $\gamma^i_{jk}$ using
\figref{BendingSchematic},
\begin{equation}
  \begin{aligned}
    \partial_1e_1 &\sim \frac{a}{(l_0/2)^2}e_3, \\
    \partial_2e_1 = \partial_1e_2 &\sim 0, \\
    \partial_2e_2 &\sim \frac{1}{a}e_3.
  \end{aligned}
\end{equation}
From this we see that the changes in the basis vectors are primarily in the
$e_3$ direction, meaning that they do not contribute to $\mathsf H_{ij}$.

If we take $l_0$ to be much larger than $a$, then the dominant term in $\mathsf
H_{ij}$ will be $1/a$, but even if $l_0$ and $a$ are a similar order of
magnitude, all of the $\mathsf H_{ij}$ terms will be of the order $1/a$. Hence,
we expect $\tr(\mathsf H^2)$ and $\tr(\mathsf H)^2$ will scale as
$(1/a)^2=(1/\SI{3}{\mm})^2=\SI{0.1}{\per\square\mm}$.

Given these scalings for $\tr(\mathsf E^2)$, $\tr(\mathsf E)^2$, $\tr(\mathsf
H^2)$, and $\tr(\mathsf H)^2$, and the values $E=\SI{1}{\MPa}$, $\nu=0.5$, and
$h=\SI{0.3}{\mm}$, we can obtain scalings for the bending and stretching energy
given in \eqref{DeformationEnergy},
\begin{equation}
  \begin{aligned}
    \text{stretching energy}&\sim\SI{0.02}{\N\per\mm}, \\
    \text{bending energy}&\sim\SI{1.5e-4}{\N\per\mm},
  \end{aligned}
\end{equation}
This indicates that given the kind of deformation we have observed in our
Starling resistors at onset, i.e. where the strain energy is dominated by the
effects of pre-strain, the axis aligned with the largest strain remains close to
perpendicular to the axis aligned with the largest curvature, rotations are
mostly about axes tangential to the shell, and changes in the rotation scale
with the change of rotation about the longitudinal axis in the azimuthal
direction, stretching energy will dominate bending energy. Moreover, this result
remains valid even when the tube length gets close to the tube diameter. This is
significant for our considerations of the lung, since the length to diameter
ratio of tubes in the lung typically varies from \num{1} to \num{6}
\citep{Horsefield:1968tc}.

%-------------------------------------------------------------------------------
\subsection{Direct Calculation from Data}
\begin{figure}
  \centering
  \includegraphics[width=13cm]{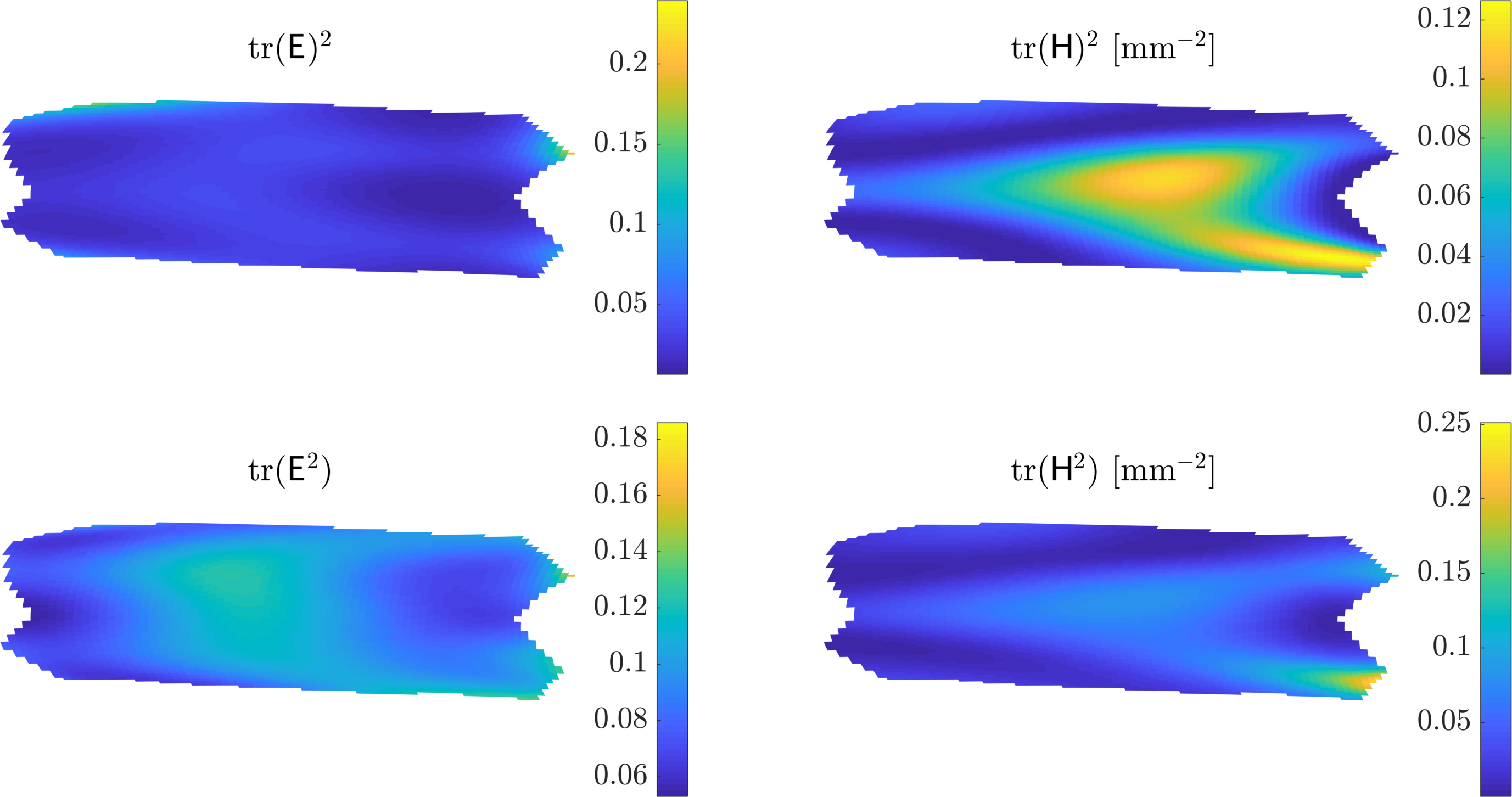}
  \caption{Values of the kinematic variables $\tr(\mathsf E^2)$, $\tr(\mathsf
    E)^2$, $\tr(\mathsf H^2)$, and $\tr(\mathsf H)^2$.}
  \label{HSV_Tre_and_Trh}
\end{figure}

\begin{figure}
  \centering
  \includegraphics[width=13cm]{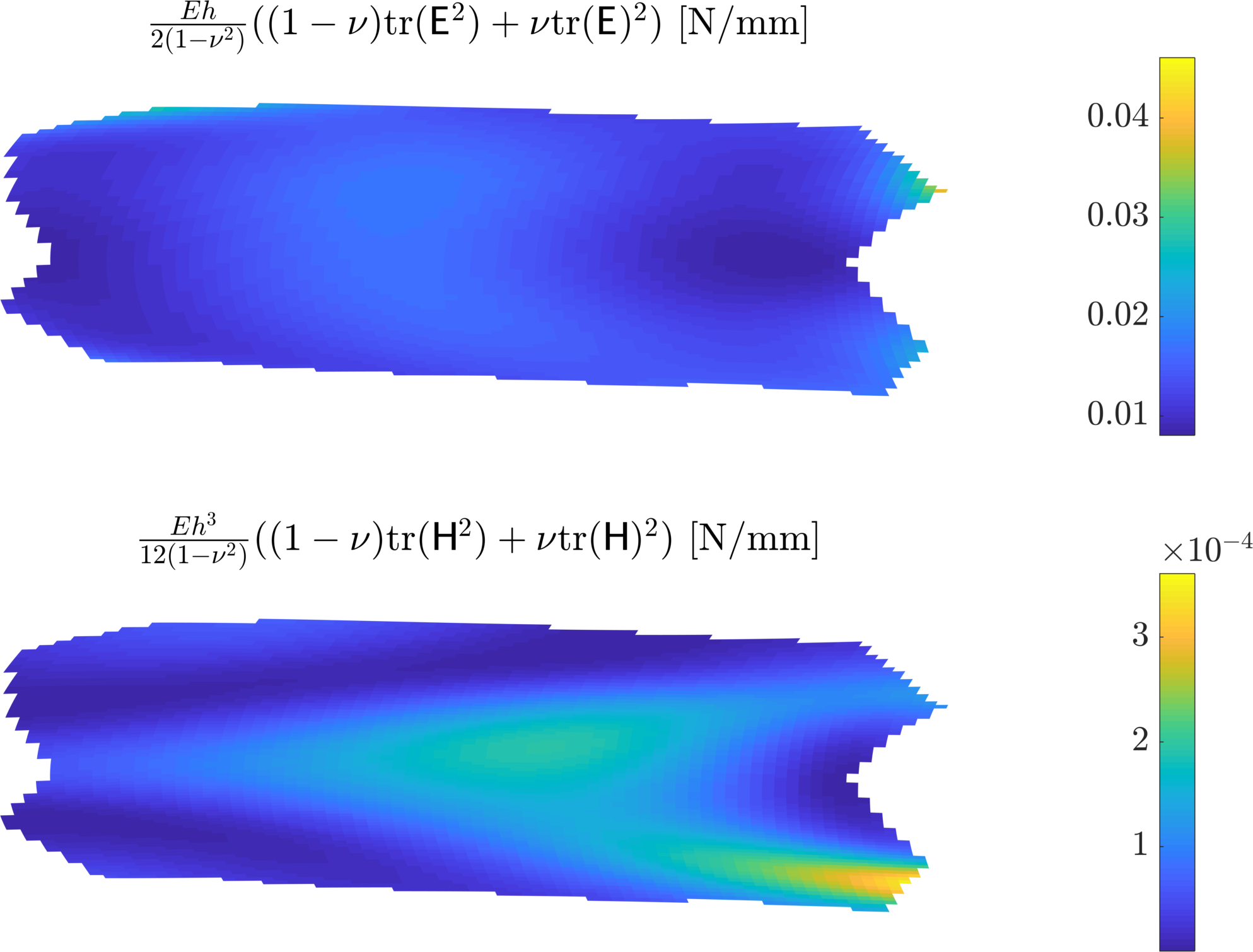}
  \caption{The stretching energy (top image) and bending energy (bottom image)
    associated with the deformation of the flexible tube.}
  \label{HSV_Vs_and_Vb}
\end{figure}

We can use the high speed video data to calculate $\tr(\mathsf E^2)$,
$\tr(\mathsf E)^2$, $\tr(\mathsf H^2)$, and $\tr(\mathsf H)^2$ from
\eqref{DeformationEnergy}. Typical plots are shown in \figref{HSV_Tre_and_Trh},
from which we see that in the units chosen these have similar orders of
magnitude. We also see that the scalings obtained in the previous section agree
with these plots well, providing support for the assumptions made. We can also
calculate the bending and stretching energy, and this is shown in
\figref{HSV_Vs_and_Vb}. This agrees very well with the scaling values of the
previous section, again supporting our conclusions.

%-------------------------------------------------------------------------------
\section{Conclusions}
We have developed a new method of representing the change of curvature tensor
using rotors (see \eqref{HRepresentation}), increasing our understanding of
bending in shells. We have used this representation to explain results from
stereographic imaging of Starling resistors that demonstrate that the bending
energy in these deformations is around 2 orders of magnitude lower than the
stretching energy.  We have been able to show that this relies on the fact that
the strain energy is dominated by the effects of pre-strain, the axis aligned
with the largest strain remains close to perpendicular to the axis aligned with
the largest curvature, rotations are mostly about axes tangential to the shell,
and changes in the rotation scale with the change of rotation about the
longitudinal axis in the azimuthal direction. Further to this, our scaling
analysis remains valid even when the tube length gets close to the tube
diameter. This is of significance to our work in understanding wheezing, since
the length to diameter ratio of tubes in the lung typically varies from \num{1}
to \num{6}. Hence we have provided a scaling analysis, confirmed by experiment,
that allows us to say that bending energy is dominated by stretching energy
during self excited oscillations in the airways of the lung. This should allow
the use of membrane theory to model the tube, which reduces the order of the
equations of motion from \num{4} to \num{2}.

\vskip1pc

\paragraph{Ethics} This paper adheres to the Royal Society publishing and
research ethics policy.

\paragraph{Data Accessibility} Data for this paper can be accessed at
\\\texttt{https://doi.org/10.17863/CAM.10363}.

\paragraph{Author Contributions} The work was completed as part of the PhD
project of AG, with AA supervising, and JL advising, in particular on image
processing and the uses of geometric algebra.

\paragraph{Competing Interests} The authors declare no competing interests.

\paragraph{Funding} The authors would like to acknowledge funding from the
EPSRC, the IMechE Postgraduate Research Scholarship, and Engineering for
Clinical Practice (\texttt{http://divf.eng. cam.ac.uk/ecp/Main/EcpResearch}).

\paragraph{Acknowledgements} Our lab technician John Hazelwood, who has been
instrumental in the experimental work completed, and Holger Babinsky, who
provided advice on the use of high speed cameras.

%%%%%%%%%% Bibliography %%%%%%%%%%%%%%
\bibliographystyle{rspub_unsrt}
\bibliography{references}
\end{document}